\documentclass[twocolumn]{revtex4}
\usepackage[dvipdf]{graphicx}
\usepackage{amssymb}
\usepackage{amsmath}
\usepackage{booktabs}
\usepackage{lscape}
\usepackage{txfonts}

\usepackage{color}

\begin{document}

\title{A stochastic reorganizational bath model for electronic energy transfer}

\author{Takatoshi Fujita}
\affiliation{Department of Chemistry and Chemical Biology, Harvard University, Cambridge, Massachusetts 02138, USA}
\email{tfujita@fas.harvard.edu;aspuru@chemistry.harvard.edu}
\author{Joonsuk Huh}
\affiliation{Department of Chemistry and Chemical Biology, Harvard University, Cambridge, Massachusetts 02138, USA}
\author{Al\'{a}n Aspuru-Guzik}
\affiliation{Department of Chemistry and Chemical Biology, Harvard University, Cambridge, Massachusetts 02138, USA}
%\email{aspuru@chemistry.harvard.edu}

\begin{abstract} Environmentally induced fluctuations of the optical gap play a crucial role in electronic energy transfer dynamics.
One of the simplest approaches to incorporate such fluctuations in energy transfer dynamics is the well known Haken-Strobl-Reineker (HSR) model, in which the energy-gap fluctuation is approximated as white noise. Recently, several groups have employed molecular dynamics simulations and excited-state calculations in conjunction to account for excitation energies' thermal fluctuations. On the other hand, since the original work of HSR, many groups have employed stochastic models to simulate the same transfer dynamics. Here, we discuss a rigorous connection between the stochastic and the atomistic bath models. If the phonon bath is treated classically, time evolution of the exciton-phonon system can be described by Ehrenfest dynamics. To establish the relationship between the stochastic and atomistic bath models, we employ a projection operator technique to derive the generalized Langevin equations for the energy-gap fluctuations. The stochastic bath model can be obtained as an approximation of the atomistic Ehrenfest equations via the generalized Langevin approach. Based on this connection, we propose a novel scheme to take account of reorganization effects within the framework of stochastic models. The proposed scheme provides a better description of the population dynamics especially in the regime of strong exciton-phonon coupling. Finally, we discuss the effect of the bath reorganization in the absorption and fluorescence spectra of ideal J-aggregates in terms of the Stokes shifts. We find a simple expression that relates the reorganization contribution to the Stokes shifts -- the reorganization shift -- to the ideal or non-ideal exciton delocalization in a J-aggregate.
The reorganization shift can be described by three parameters: the monomer reorganization energy, the relaxation time of the optical gap, and the exciton delocalization length. This simple relationship allows one to understand the physical origin of the Stokes shifts in molecular aggregates.
\end{abstract}

\maketitle

\section{INTRODUCTION}
%atural and artificial light-harvesting systems, molecular aggregates.
Natural photosynthesis starts with light energy absorption by an assembly of photosynthetic pigments, after which this excitation energy is transferred to a reaction center,~\cite{Blankenship2002,Valkunas1996,Grondelle2006,Engel2007,Caruso2009,Cheng2009,Panitchayangkoon2010,Scholes2011,Nalbach2013} where charge transfer is carried out.
In most organisms, various pigment-protein complexes are responsible for this light harvesting. 
The highly-ordered structures of light-harvesting complexes have motivated researchers to design artificial light-harvesting antenna systems such as self-assembled supramolecular systems~\cite{Li2004,Balaban2005,Wasielewski2009}, quantum dots~\cite{Robel2006,Xin2010}, and metal-organic frameworks~\cite{Wang2012,Son2013}.
%The artificial light-harvesting systems, such as J-aggreagates~\cite{Wurthner2011}, conjugated polymer~\cite{Collini2009}, and metal-organic framework.
%In natural light-harvesting systems, pigment molecules molecular aggregates~\cite{Ishizaki2010,Saikin2013}
%J-aggregates refs. Wurthner, Eisele, Balaban, Tamiaki
% MOF, DNA-templated dye
% excitonics

% theoretical treatment of EET, open quantum system approaches, brief view of theoretical treatment
Experimental studies of natural and artificial light-harvesting systems have spurred theoretical descriptions of electronic energy transfer (EET) in complex systems. EET has been modeled using F\"{o}rster theory~\cite{Forster1959,Ern1972,Rieger1997} which describes exciton transport as incoherent hopping between chromophores. However, this approach is applicable only in the strong exciton-phonon coupling regime, where the excitonic couplings between chromophores are small relative to the exciton-phonon couplings.
In the opposite limit, a quantum master equation can be derived by treating the exciton-phonon coupling perturbatively~\cite{May2011}. The most commonly used theory from this limit is the Redfield approach~\cite{Redfield1965}. In order to more accurately model EET in the intermediate regime between these two limiting cases, one could consider using the hierarchy equations of motions (HEOM)~\cite{Tanimura1989,Ishizaki2009a,Ishizaki2009b,Rebentrost2011,Zhu2011,Kreisbeck2011,Kreisbeck2012,Aghtar2012,Arend2012}, path integral Monte Carlo~\cite{Muhlbacher2004,Muhlbacher2012}, a polaron-transformed master equation~\cite{Jang2008,Jang2011,Kolli2011}, non-Markovian quantum state diffusion~\cite{Diosi1998,Ritschel2011a,Ritschel2011b}, or density matrix renormalization group methods~\cite{Chin2011}.
Different methodologies for simulating energy transfer have been widely reviewed~\cite{Makri1999,Ishizaki2010,Kuhn2011,Pachon2012,Biele2012,Renger2013,Saikin2013,Jang2013}. The most expensive and accurate methods are unified models that combine electronic structure directly with bath models~\cite{Parkhill2012}.

% correlation function, spectral density, stochastic models vs atomistic bath models 
% role of structured phonon bath
The details of exciton-phonon coupling are crucial in light-harvesting systems because they strongly modify the excited-state dynamics.
The environmental effect is usually characterized as a two-time bath correlation function or a bath spectral density. 
The Haken-Strobl-Reinker (HSR)~\cite{Haken1972,Haken1973} model approximates bath fluctuation as white noise.
The model can be extended to treat colored noise~\cite{Warns1998,Barvik1999}. 
However, HSR-like approaches are incapable of describing bath reorganization and finite temperature effects.
A lack of a bath reorganization process causes the dephasing rate to be underestimated~\cite{May2011} and leads to the same peak positions for absorption and fluorescence spectra without the Stokes shift.
%~\cite{Haken1972,Haken1973,Rebentrost2009a,Valleau2012} 
% or Kubo-Anderson~\cite{Anderson1954,Kubo1954,Fujita2012} models.
As mentioned previously, in recent years, exciton-phonon interactions have been treated at the atomistic level by combining molecular dynamics simulations and excited-state calculations~\cite{Damjanovic2002,Olbrich2011a,Olbrich2011b,Olbrich2011c,Shim2012,Fujita2012}.
From these simulations, one can perform ensemble-averaged wavepacket dynamics~\cite{Ishizaki2011,Zhong2011,Berkelbach2012,Aghtar2012}, also called Ehrenfest dynamics, or extract a spectral density~\cite{Valleau2012b} for its use in a quantum master equation.

In earlier work on the Fenna-Matthes-Olson complex~\cite{Shim2012}, we compared the atomistic Ehrenfest dynamics -- the atomistic bath model -- with the stochastic HSR model. 
Surprisingly, these two methods gave similar exciton population dynamics regardless of the different descriptions of the phonon baths. 
Motivated by these results, we discuss a rigorous connection between the atomistic and stochastic bath models and develop a novel correction scheme for the stochastic approaches. Atomistic or parameterized Ehrenfest dynamics require the time evolution of many bath degrees of freedom. The advantage of the stochastic approaches lies in the use of collective bath variables. This computational simplicity of the stochastic bath model will be useful to simulate EET in large systems, for example, biological scales.
To relate the atomistic and stochastic bath models, we exploit the projection operator technique~\cite{Mori1965} to derive the generalized Langevin equations for energy-gap fluctuations. 
The stochastic model can be derived directly from the atomistic Ehrenfest equations via the generalized Langevin approach.
Based on the connection between the stochastic and the atomic approaches, we propose a novel scheme to correct a reorganization effect in the stochastic model. Finally, our scheme will be compared with the HEOM through the population dynamics of a model system. 

%organization of paper
The remainder of this manuscript is organized as follows: 
In Sec. II, we introduce the exciton-phonon Hamiltonian and the Ehrenfest equations. Next, we exploit the projection operator technique in the atomistic Ehrenfest equations to derive stochastic bath models. Furthermore, we propose a scheme to correct reorganization effects within the stochastic bath models. 
In Sec. III, we present numerical results for the population dynamics and Stokes shifts in idealized J-aggregates.
In Sec. IV, we present the manuscript with concluding remarks.

\section{THEORY}
\subsection{Excitonic Hamiltonian and Ehrenfest equations}
% Hamiltonian
The excitonic Hamiltonian in the single exciton manifold of a molecular aggregate can be written as follows~\cite{May2011}: 
\begin{equation}
H= \sum_m^{N} \epsilon_m(\mathbf{R}) \vert m\rangle \langle m\vert 
+ \sum_{m\neq n}^{N}V_{mn}(\mathbf{R})\vert m \rangle \langle n\vert + T(\mathbf{P}) + V_{G}(\mathbf{R}),
\end{equation}
where $\vert m\rangle$ denotes the state where an electronic excitation is localized at $m$-th molecule (site) and all other molecules are in the ground states. $\mathbf{R}$ and $\mathbf{P}$ refer to the nuclear coordinates and momentum, respectively.
$\epsilon_m(\mathbf{R})$ represents the excitation energy of $m$-th site in the nuclear configuration $\mathbf{R}$, and $V_{mn}(\mathbf{R})$ is an excitonic coupling constant between $m$-th and $n$-th molecules. Here, $T(\mathbf{P})$ and  $V_G(\mathbf{R})$ are the kinetic energy and the ground-state potential energy for the nuclear coordinates. Hereafter, the dependence of $V_{mn}$ on $\mathbf{R}$ is neglected. Then, we decompose the total Hamiltonian into the form of the system-bath Hamiltonian as follows:
\begin{equation}
H_S=\sum_m^N \left< \epsilon_m \right >\vert m \rangle \langle m\vert  + 
\sum^N_{m \neq n}  V_{mn}\vert m \rangle \langle n\vert,
\end{equation}
\begin{equation}
H_{SB}=\sum_m^N \Delta \epsilon_m(\mathbf{R})\vert m\rangle \langle m\vert,
\end{equation}
\begin{equation}
H_{B}=T(\mathbf{P})+V_G(\mathbf{R}).
\end{equation}
Here, $\Delta \epsilon_m(\mathbf{R})=\epsilon_m(\mathbf{R}) -  \left< \epsilon_m\right >$, and a bracket $\left<  \right>$ denotes the ensemble average over $H_{B}$.

If we treat the nuclear degree of freedoms classically, the wave function of the exciton system is described with the time-dependent Schr$\ddot{\rm{o}}$dinger equation,
\begin{equation}
\label{eqn:sch}
i\hbar \frac{\partial}{\partial t}\vert \psi(t) \rangle=\left( H_S+H_{SB}(t) \right) \vert \psi(t) \rangle.
\end{equation}
The nuclear degrees of freedom follow from Hamilton's equation in the mean-field interaction, 
\begin{equation}
\dot{\mathbf{R}}=\frac{\partial H_{E}}{\partial \mathbf{P}},
\end{equation}
\begin{equation}
\dot{\mathbf{P}}=-\frac{\partial H_{E}}{\partial \mathbf{R}},
\end{equation}
with the Hamiltonian for the nuclei on the excited-state potential energy surface (PES):
\begin{align} 
\label{eqn:He}
H_{E}=\left<\psi(t)\left|H \right|\psi(t)\right> \\ \notag
	=H_B + H'(t),
\end{align}
\begin{equation}
H'(t)=\sum_m^N \Delta \epsilon_m(\mathbf{R})\left|\left<m|\psi(t)\right>\right|^2.
\end{equation}

In Eq.~\ref{eqn:He}, $H'(t)$ gives the Ehrenfest mean potential; that is, the average of site energies weighted by exciton populations, $\left|\left<m|\psi(t)\right>\right|^2$. This term shifts the excited-state PES with respect to the ground-state one, as illustrated in Fig. 1. The PES shift is essential to describe reorganization processes after the photoexcitation. This effect is also referred to as a back reaction~\cite{Aghtar2012}, because through this term the dynamics of the exciton system can affect the dynamics of the phonon bath.

\begin{figure}[!ht]
\begin{center}
 	\includegraphics[width=8cm]{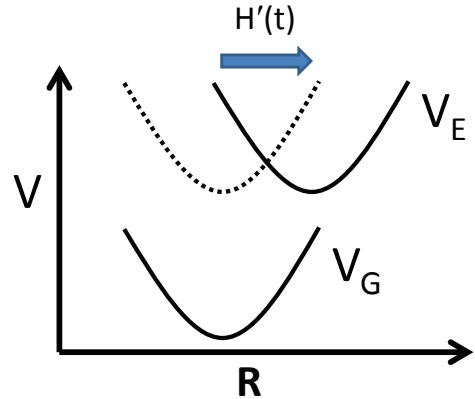}
\end{center}
\caption{Schematic of $H'(t)$, which shifts the excited-state PES compared with the ground-state one. $V_E$ refers to the excited-state PES: $V_E$=$V_G$+$H'(t)$}
\end{figure}

The density matrix of the excitonic system $\rho$ is obtained as the average of an ensemble of unitary evolutions:
\begin{equation}
\label{eqn:ens}
\rho(t)=\frac{1}{M}\sum_i^M \left| \psi_i(t) \rangle \langle \psi_i(t) \right|.
\end{equation}

Eqs. 5-10 form the basis for mixed quantum-classical approaches for exciton dynamics~\cite{Damjanovic2002,Olbrich2011a,Olbrich2011b,Olbrich2011c,Shim2012}.
In these applications, the Ehrenfest mean force (i.i., the second term in Eq.~\ref{eqn:He}) is neglected, because one needs to run MD simulations in excited-state PES, which is computationally more expensive to determine. This term may be ignored when the reorganization energy is small relative to the excitonic couplings. Later, we discuss the condition under which this term can be ignored.

\subsection{Generalized Langevin equations for energy-gap fluctuations}
In this section, we derive stochastic bath models from the atomistic Ehrenfest equations.
To relate the atomistic and stochastic bath models, we apply the projection operator technique~\cite{Zwanzig1961,Mori1965,Berne1970} first developed in classical statistical mechanics.
The projection operator techniques can provide a microscopic derivation of a phenomenological equation such as the Langevin equation. 
%We specifically apply Mori's projection operator technique~\cite{Mori1965} to the exciton-phonon system in order to derive generalized Langevin equations for the energy-gap fluctuations.
We briefly review Mori's projection operator formalism~\cite{Mori1965} and the derivation of the generalized Langevin equations. 
The standard technique is applied to the exciton-phonon system to derive stochastic equations for the energy-gap fluctuations.

%classical Lioville operators
Here, we consider the time evolution of the site energy fluctuations,
\begin{equation}
%\[ 
\Delta \boldsymbol{\epsilon}(t) = \left(
\begin{array}{c}
\Delta \epsilon_1(t) \\
\vdots \\
\Delta \epsilon_N(t) 
\end{array}
\right).
% \]
\end{equation}
In classical mechanics, any physical variable $A$ is a function of $\mathbf{R}$ and $\mathbf{P}$, and the time evolution of $A$ is given by a Poisson bracket with the Hamiltonian. Accordingly, the equations of motion of $\Delta \boldsymbol{\epsilon}(t)$ are given by, 
\begin{align}
\frac{\partial \Delta \boldsymbol{\epsilon}}{\partial t} & =\left\{\Delta \boldsymbol{\epsilon}, H_{E} \right\} \\ \notag
	&= iL_0\Delta \boldsymbol{\epsilon} + iL'\Delta \boldsymbol{\epsilon}.
\end{align}
Here, we introduce the Liouville operators of $H_B$ and $H'(t)$ as follows:
\begin{equation}
iL_0=\left\{~~~, H_B\right\},
\end{equation}
\begin{equation}
iL'(t)=\left\{ ~~~, H'(t)\right\}.
\end{equation}

% generalized Langevin equations
First, we neglect the $iL'$ term in such a way that the time evolution of $\Delta \boldsymbol{\epsilon}$ can be determined solely from $iL_0$. This approximation leads to the exciton dynamics ignoring the reorganization effects.
We then apply the projection operator technique~\cite{Mori1965} to derive the generalized Langevin equations for $\Delta \boldsymbol{\epsilon}$.
We define the projection operator $P$ from the site-energy fluctuations:
\begin{equation}
PA(t)=\left(A(t),\Delta \boldsymbol{\epsilon} \right)\left(\Delta \boldsymbol{\epsilon},\Delta \boldsymbol{\epsilon}\right)^{-1}\Delta \boldsymbol{\epsilon}.
\end{equation}
In this paper, we will use a simplified notation $(A,B) \equiv \left<BA^*\right>$.
By applying the projection operator to Eq. 12, the time evolution for $\Delta \boldsymbol{\epsilon}(t)$ is obtained as the generalized Langevin equation
\begin{equation}
\label{eqn:glang}
\frac{\partial}{\partial t}\Delta \boldsymbol{\epsilon}(t)=-\int^t_0ds\boldsymbol{M}(t-s)\cdot\Delta\boldsymbol{\epsilon}(s)+\boldsymbol{F}(t).
\end{equation}
Here, the memory matrix $\boldsymbol{M}(t)$ and the random force vector $\boldsymbol{F}(t)$ are defined as
\begin{equation}
\boldsymbol{F}(t)= \exp{(iQLQt)}\Delta \dot{\boldsymbol{\epsilon}},
\end{equation}
\begin{equation}
\boldsymbol{M}(t)= \left(\boldsymbol{F},\boldsymbol{F}(t) \right)\left(\Delta\boldsymbol{\epsilon},\Delta \boldsymbol{\epsilon}\right)^{-1},
\end{equation}
where $Q=1-P$.

%derivation of stochastic models
Given the generalized Langevin equations for the site-energy fluctuations, deriving stochastic models for them is straightforward~\cite{Anderson1954,Kubo1954,Haken1972,Haken1973}.
We introduce the Markov approximation to the memory functions and neglect cross correlations between different sites as follows:
\begin{equation}
\label{eqn:markov}
M_{mn}(t)= 2\Gamma_m \delta_{mn}\delta (t).
\end{equation}
By using this approximation, Eq.~\ref{eqn:glang} become a set of Langevin equations with Gaussian fluctuations $F_m(t)$,
\begin{equation}
\label{eqn:lang}
\frac{\partial}{\partial t}\Delta \epsilon_m(t)=-\Gamma_m\Delta \epsilon(t) +F_m(t),
\end{equation}
where $\left <F_m(t) \right>=0$ and $\left < F_m(t)F_m(0)\right>=2\left<\Delta \epsilon_m^2\right>\Gamma_m\delta(t)$.
$\Gamma_m$ is defined as 
\begin{equation}
\frac{1}{\Gamma_m}=\int_0^{\infty}dt\frac{\left <\Delta \epsilon_m(t) \Delta \epsilon_m(0)\right>}{\left<\Delta \epsilon_m^2 \right>}.
\end{equation}
The inverse of $\Gamma_m$ is related to the decay time $1/\Gamma_m = \tau_m$, which characterizes the relaxation time of the bath fluctuations.
As the correlation function induced by Eq.~\ref{eqn:lang} is an exponentially decaying function, i.e., 
\begin{equation}
\label{eqn:cexp}
\left < \Delta \epsilon_m(t)\Delta \epsilon_m(0)\right>=\left<\Delta \epsilon_m^2 \right>e^{-\Gamma_mt},
\end{equation}
this method is identical to the Kubo-Anderson (KA) stochastic model~\cite{Anderson1954,Kubo1954}.

Furthermore, if we consider the limit of $\tau$$\to$0, the correlation function becomes a delta function:
\begin{equation}
\left < \Delta \epsilon_m(t)\Delta \epsilon_m(0)\right>=\frac{2\left<\Delta \epsilon_m^2 \right>\delta (t)}{\Gamma_m}.
\end{equation}
%So far, we have shown that the stochastic bath models, such as the KA or HSR models, are the limiting case of the atomistic Ehrenfest equations.
So far, we have discussed that the stochastic bath models, such as the KA or HSR models, are the limiting case of the atomistic Ehrenfest equations.
The stochastic bath models are known to result in high-temperature dynamics and are incapable of describing reorganization processes. The high-temperature dynamics results from the classical treatment of the phonon bath. Classical correlation functions do not satisfy the detailed balance condition. Within the stochastic Schr\"{o}dinger equations, the non-Markovian terms and the detailed balance condition are important for describing the thermal relaxation process~\cite{Biele2012}.
The lack of reorganization is due to the neglect of the $H'(t)$ term, which plays an essential role in the reorganization process that follows photoexcitation.

%advantage of stochastic method. inhomogeneous vs homogeneous system
The parameters $\Gamma_m$ and $\left<\Delta \epsilon_m^2 \right>$ for the dynamical equations can be obtained from MD simulations and excited-state calculations. 
The proposed simplified descriptions of the bath fluctuations allow us to simulate EET in large systems. 
For example, in earlier works, we used the KA and HSR approaches to simulate the light-harvesting apparatus of green-sulfur bacteria, which consists of thousands of chromophores~\cite{Fujita2012,Huh2014,Fujita2014}.

% correlated fluctuation
%A possible extension is the cross-correlation of the memory functions.
%In this case, we approximate the memory function as 
%Treatment of non-Markov effect of the memory function may be achieved by the continued fractal expansion~\cite{Mori1965b,DeRaedt1977}.

%In this case we approximate the memory matrix as follows:
%The site energy fluctuations are coupled each other.
%\begin{equation}
%\boldsymbol{\Gamma} =\left <\Delta \epsilon_m(t) \Delta \epsilon_m(0)\right>\left<\Delta \epsilon \right>.
%\end{equation}
%where $\left <\boldsymbol{F}_m(t) \right>=0$ and $\left < \boldsymbol{F}_m(t)\boldsymbol{F}_m(0)\right>=2\left<\Delta \epsilon_m^2\right>\delta(t)/\tau_m$.
%This method can be an alternative method to examine the correlated fluctuation in photosynthetic system.
%The parameters for cross-correlations have been treated as phenomenological parameters~\cite{Rebentrost2009c,Dijkstra2010,Sebastiaan2012}

\subsection{Reorganization correction to stochastic bath model}

%Langevin equations with purturbation potentials
In this section, we modify the KA model such that the reorganization effect can be included via the Ehrenfest mean potential.
We begin by the exact time evolution of the site-energy fluctuation in the excited-state manifold.
The dynamics $\Delta \boldsymbol{\epsilon}$ in the excited-state can be given by the time-evolution operator, $e^{\int_0^tds\left(iL_0+iL'(s)\right)}$,
\begin{equation}
\frac{\partial}{\partial t}\Delta \boldsymbol{\epsilon}(t)=\frac{\partial}{\partial t}e^{\int_0^tds\left(iL_0+iL'(s)\right)}\Delta \boldsymbol{\epsilon}(0).
\end{equation}
The definition of $\Delta \epsilon_m$ has not been changed, i.e. $\Delta \epsilon_m = \epsilon - \left< \epsilon \right>_{g}$.
Here, $\left<\right>_{g}$ and $\left< \right>_{e}$ refer to the ensemble average over ground- and excited-state PESs, respectively.
The first-order expansion of the time-evolution operator with respect to $iL'(t)$ gives
\begin{equation}
\frac{\partial}{\partial t}\Delta \boldsymbol{\epsilon}(t)=\frac{\partial}{\partial t}e^{iL_0t}\Delta \boldsymbol{\epsilon}(0)+\frac{\partial}{\partial t}\int_0^tdse^{iL_0(t-s)}iL'(s)e^{iL_0s}\Delta \boldsymbol{\epsilon}(0).
\end{equation}
The first term in the right-hand side is identical to Eq.~\ref{eqn:glang},.i.e. the time evolution in the ground-state. The second term in the right-hand side is the first-order correction of the reorganization effects. We redefine the time-local second term by $\boldsymbol{K}(t)$ that is a characteristic force induced by $H'(t)$. Finally, we can obtain the following equation for $\Delta \boldsymbol{\epsilon}$:
\begin{equation}
\frac{\partial}{\partial t}\boldsymbol{\Delta \epsilon}(t) = -\int^t_0ds\boldsymbol{M}(t-s)\cdot\Delta\boldsymbol{\epsilon}(s)+\boldsymbol{F}(t) + \boldsymbol{K}(t).
\end{equation}
Within the first-order perturbation with respect to $iL'$, we can correct the reorganization effects by adding the time-local term to the Langevin equations.
After applying the same approximation as Eq.~\ref{eqn:markov}, the time evolution of the $\Delta \boldsymbol{\epsilon}$ in the excited-state follows the Langevin equation with the white noise,

%We attempt to describe the dynamics of $\Delta \boldsymbol{\epsilon}$ in the excited-state PES with the following Langevin equation:
\begin{equation}
\label{eqn:lane}
\frac{\partial}{\partial t}\Delta \epsilon_m = -\Gamma_m \Delta \epsilon_m(t) + F_m(t) + K_m(t).
\end{equation}

The formal solution of Eq.~\ref{eqn:lane} is
\begin{equation}
\label{eqn:slane}
\Delta \epsilon_m(t) = e^{-\Gamma_m t}\Delta \epsilon_m(0)+\int_0^t ds e^{-\Gamma_m(t-s)}\left( F_m(s)+K_m(s)\right).
\end{equation}
After taking the ensemble average of Eq.~\ref{eqn:lane} and using the definition of $\Delta \epsilon$ and $F(t)$, we have
\begin{align}
\label{eqn:ss}
\left< \Delta \epsilon_m(t) \right>                             & = 
\left< \epsilon_m(t)\right>_e-\left< \epsilon_m\right>_g & =\int_0^tdse^{-\Gamma_m(t-s)}K_m(s).
\end{align}
Here, we consider a situation where the excitonic system is excited at $t$ = 0, resulting in the same nuclear configuration which implies that $\left< \epsilon_m(0)\right>_e=\left< \epsilon_m\right>_g $
Intuitively, $K_m(t)$ describes the Stokes shifts and thus the reorganization process.

% Identify K(t) by linear-response theory
In the following, we use the linear-response theory to obtain $K_m(t)$.
The first-order perturbation of site energies due to $H'(t)$~\cite{Kubo1957} is
\begin{equation}
\left< \epsilon_m(t)\right>_e-\left< \epsilon_m\right>_g= \beta \sum_n \int^t_0ds\left<\Delta \dot{\epsilon}_m(t-s)\Delta \epsilon_n(0)\right>P_n(s),
\end{equation}
where $P_n(t) = \left<n\left|\rho\right|n\right>$, and the inverse temperature is denoted by $\beta$ = 1/($k_BT$).
We approximate the correlation function as an exponential (Eq.~\ref{eqn:cexp}) in such a way that it coincides with the previous derivations.
Thus, the difference in average site energies is given by
\begin{equation}
\left< \epsilon_m(t)\right>_e-\left< \epsilon_m\right>_g=-\beta\Gamma_m\left< \Delta\epsilon_m^2\right>\int_0^tdse^{-\Gamma_m(t-s)}P_m(s).
\end{equation}
By comparing this result with Eq.~\ref{eqn:ss}, we identify $K_m(t)$ as
\begin{equation}
K_m(t)=-\beta\Gamma_m\left< \Delta \epsilon_m^2\right>_gP_m(t).
\end{equation}
This is the main result leading to a new term in the stochastic equations and the description of the reorganization effects. The steady-state solution of Eq.~\ref{eqn:lane} is $\left< \epsilon_m \right>_e-\left< \epsilon_m\right>_g=-\beta \left< \Delta\epsilon_m^2\right>P_m$. 
Note that, in the high-temperature limit, the variance and reorganization energy $\lambda$ are related by $\left<\Delta \epsilon^2 \right>_g=2k_BT\lambda$.
If a system is composed of a single pigment ($P$=1), then it is easy to see that one gets $\left< \epsilon\right>_e-\left< \epsilon\right>_g = -2\lambda$.
This is a well-known result that the Stokes shift is twice the reorganization energy. As expected, $K_m(t)$ shifts the excited-state PES and induces the Stokes shifts. In Fig. 2, we use Eqs. 20 and 24 to show the distribution of site energies for the ground and excited-state of a two-level molecule. As expected, the distributions are shifted by $2\lambda$.

By solving Eq.~\ref{eqn:sch} with Eq. \ref{eqn:lane}, we have shown that the reorganization effects can be incorporated within the framework of a stochastic bath model. For the remainder of this paper, we refer to our scheme as the reorganization-corrected Kubo-Anderson (RECKA) model.
The exciton populations $P_m$ are necessary for input into RECKA, and they can be obtained after taking an ensemble average of unitary evolutions. Therefore, Eqs.~\ref{eqn:sch}, \ref{eqn:lane}, and \ref{eqn:ens} must be solved self-consistently until $P_m$ converge.

\begin{figure}[!ht]
\begin{center}
 	\includegraphics[width=6cm]{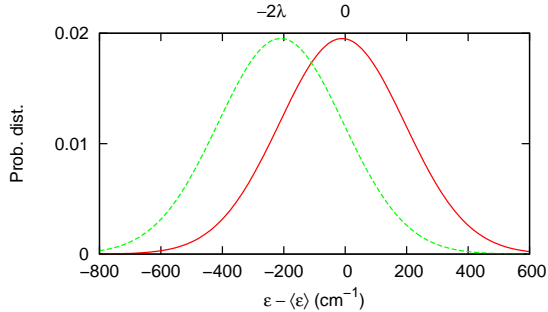}
\end{center}
\label{fig:dist}
\caption{Probability distributions of site energy on ground (red-solid) and excited (green-dashed) potential energy surfaces, which were obtained from the Langevin equation without (Eq.~\ref{eqn:lang}) or with (Eq.~\ref{eqn:lane}) the reorganization correction. Here, the Langevin equations were solved with $T$= 300 $K$ and $\Gamma$ = $\lambda$ = 100 cm$^{-1}$}
\end{figure}
We note that the newly proposed term, $K_m(t)$, can be determined by the variance and the inverse of relaxation times, so additional computational time is not required.
The lack of additional computational expense is a huge advantage of the RECKA over others methods that evaluate Ehrenfest potentials accurately; these methods demand the derivative calculation of the excited-state PES, which is a computationally-demanding task that limits such approaches to small systems~\cite{Tully1971}.

Noticeably, the low cost of employing linear-response theory makes the applicability of the approximation reasonable for a certain range of parameters. We obtain the reorganization correction within the linear response theory. First-order perturbation provides a reasonable approximation when the Ehrenfest mean potential is small. This is true for small reorganization energy or for excitons delocalized over a large number of chromophores, $N$. The first condition is readily observed; if the reorganization energy is small, the response of the site energy is small enough to be treated by the first-order perturbation. The second condition suggests that the response becomes small for large systems irrespective of the magnitude of the reorganization energy. Because the site population is of the order $1/N$ owing to the normalization condition of the delocalized exciton populations, the response from second-order perturbation goes as $1/N^2$. This analysis indicates that the higher-order perturbations diminish with increasing $N$.

Here we mention possible generalizations of this approach. The present method can be extended to treat correlated bath fluctuations by keeping the off-diagonal terms in the memory matrix. The correlated site-energy fluctuations can be obtained by solving coupled Langevin equations. Another possible extension to include non-Markovian effects of the memory function is continued fractal expansion~\cite{Mori1965b,DeRaedt1977,Martinazzo2011}. The exponential correlation functions given by the KA or RECKA models correspond to the Lorentz-Drude spectral densities. By introducing non-delta-function memory kernels~\cite{Berne1966}, we can treat a structured spectral density and vibronic coupling~\cite{Roden2012,Chin2013}. The effect of the structured spectral density in optical and transport properties of a biological-scaled light-harvesting system is being studied in our group.~\cite{Nico2014}

\section{RESULTS AND DISCUSSION}
\subsection{Population dynamics}
%population dynamics, explanations of model and parameters
To demonstrate the reorganization scheme, we begin by considering an homogeneous dimer, i.e. $\left<\epsilon_1\right> = \left<\epsilon_2\right>$. We fix the excitonic couplings $V_{12}$=$V_{21}$ $\equiv$ $V$ to 100 cm$^{-1}$ and the temperature to 300 K. The initial conditions of site-energy fluctuations were sampled from a Gaussian distribution of variance $\left< \Delta \epsilon^2\right>$ $\equiv$ $\Delta^2$ = 2$k_BT\lambda$. Numerical results were obtained by averaging 50000 trajectories with a timestep of 1.0 fs. To examine the accuracy of our approach, we essentially compare the KA and RECKA methods with the numerically-exact HEOM approach pioneered by Tanimura and co-workers.~\cite{Tanimura1989,Ishizaki2009a,Ishizaki2009b,Rebentrost2011,Zhu2011,Kreisbeck2011,Kreisbeck2012,Aghtar2012,Arend2012} with the Lorentz-Drude spectral density of the same parameters.
We chose the inverse relaxation times $\Gamma/V$ to be 0.1, 1, and 10, which approximately correspond to 531, 53.1 and 5.3 fs, respectively. The range of suggested time depends on the system studied and the assumptions made by the authors. For example, Ishizaki and Fleming~\cite{Ishizaki2009b} proposed a slow bath correlation time greater than 100 fs for bacteriochlorophyll $a$ in the photosynthetic Fenna-Matthews-Olson complex. However, atomistic simulations have predicted much faster relaxation times~\cite{Olbrich2011b,Shim2012} of  5-10 fs. 

%population dynamics with lambda = 100cm-1 (and 500cm-1)
In Fig. 3, we compare the Ehrenfest-based methods with the HEOM for $\lambda$ = 100 cm$^{-1}$ and different relaxation times.
The population dynamics from the KA model agrees well with HEOM, which is consistent with earlier works~\cite{Ishizaki2011,Berkelbach2012,Aghtar2012}.
They show a slight deviation for the fast relaxation time of 5.3 fs.
Comparing the KA and RECKA models, we have found that the effects of reorganization corrections are negligible for slow and intermediate bath relaxation times.  However, the relaxation included in the RECKA approach for fast relaxation times yields results that are nearly identical to those of the benchmark HEOM approach.

Next, we turn to the case of a stronger exciton-phonon coupling of $\lambda$ = 500 cm$^{-1}$. The population dynamics are presented in Fig. 4 for the same series of relaxation times as those in Fig. 3. KA and RECKA still agree well with HEOM for slow and intermediate bath relaxation times. However, they deteriorate for the faster relaxation time: KA predicts a much faster time scale for inter-site energy transfer. Owing to the lack of reorganization effects, the KA approach provides the same energy-level distributions for absorption and fluorescence, which results in larger estimated hopping rates. This deficiency can be improved by the RECKA approach that exhibits a population decay similar to that of HEOM. In terms of population dynamics, we have obtained encouraging results for the set of parameters considered here.

Previous with studies using the atomistic Ehrenfest-based methods neglect the Ehrenfest mean-field potential in their description of the exciton dynamics~\cite{Damjanovic2002,Olbrich2011a,Olbrich2011b,Olbrich2011c,Shim2012}. This approximation can be justified when the reorganization energies are small compared to the excitonic couplings.
We have found that, in addition to the small reorganization energies, slow bath fluctuation is an important condition for justifying the neglect of the reorganization effect.

%limitation, Boltzmann distribution, comment on previous work by Aghtar
Similar to the case for the HSR, KA and Ehrenfest approaches, our correction scheme does not provide a Boltzmann distribution in the long-time. As discussed earlier, this deficiency originates from the classical treatment of the phonon bath and cannot be improved by reorganization correction. Based on previous work by Bastida et al.~\cite{Bastida2006}, Aghtar et al.~\cite{Aghtar2012} have introduced a correction factor that leads to the fulfillment of detailed balance. Although their scheme gives the correct thermal equilibrium distribution, it overestimates dephasing rates. An explicit quantum correction for the thermal limit may be derived by considering multiconfigurational Ehrenfest dynamics~\cite{Shalashilin2009,Shalashilin2011}, which lies outside the scope of this paper.
\begin{figure}[!ht]
\begin{center}
 	\includegraphics[width=8cm]{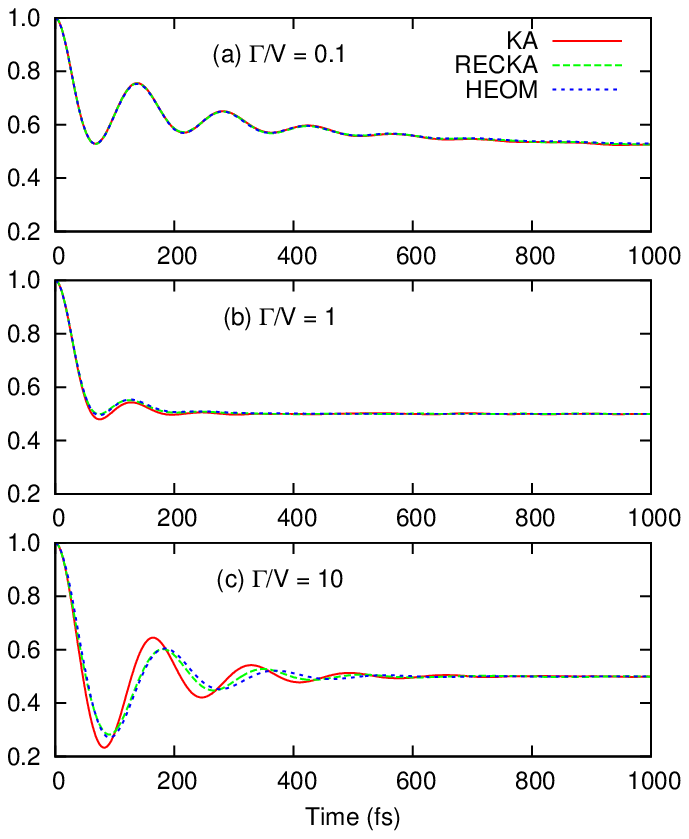}
\end{center}
\label{fig:pop1}
\caption{Populations of excited site in the dimer with equal average site energies, $V$ = 100 cm$^{-1}$, $\lambda$=100cm$^{-1}$, and $T$=300K, which were obtained rom KA (red-solid), RECKA (green-dashed), and HEOM (blue-dotted). The different relaxation times considered were (a) $\Gamma/V$=0.1 ($\tau \approx$ 530 fs), (b) $\Gamma/V$=1 ($\tau \approx$ 53 fs), and (b) $\Gamma/V$=10 ($\tau \approx$ 5.3 fs).}
\end{figure}

\begin{figure}[!ht]
\begin{center}
 	\includegraphics[width=8cm]{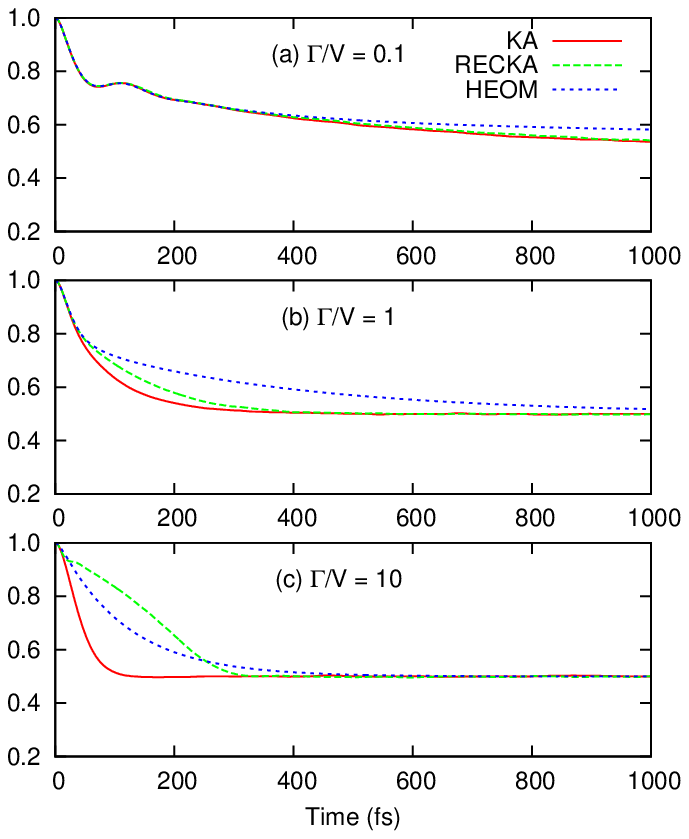}
\end{center}
\label{fig:pop2}
\caption{Populations of excited site in the dimer with equal average site energies, $V$ = 100 cm$^{-1}$, $\lambda$=500cm$^{-1}$, and $T$=300K, which were obtained from KA (red-solid), RECKA (green-dashed), and HEOM (blue-dotted). The different relaxation times considered were (a) $\Gamma/V$=0.1 ($\tau \approx$ 530 fs), (b) $\Gamma/V$=1 ($\tau \approx$ 53 fs), and (c) $\Gamma/V$=10 ($\tau \approx$ 5.3 fs).}
\end{figure}

\subsection{Stokes shifts in ideal J-aggregates}
%mini introduction. Stokes shifts in J-aggregates
J-aggregates~\cite{Jelley1936,Mobius1995,Wurthner2011}, named after Jelley, are relatively ordered molecular arrays whose transition dipoles are arranged in such a way that large exciton delocalization is present in the low-energy excitation bands. They have technologically-relevant optical properties such as a narrow absorption band that exhibits a large absorption intensity at the expense of a large number of quasi-dark excited states.
An additional optical property of J-aggregates is a small Stokes shift. 
The Stokes shifts of aggregates include internal energy relaxation among exciton states and the reorganization shifts of the exciton level. Because these two contributions cannot be separated, information on exciton-phonon coupling in aggregates cannot be obtained directly from experimental Stokes shifts.

Here, we discuss exciton-phonon coupling in ideal J-aggregates in terms of the Stokes shift magnitude.
The Hamiltonian of ideal J-aggregates is $H_S=\sum_m^N \left\{ \epsilon \vert m \rangle \langle m\vert  + V\left( \vert m \rangle \langle m+1\vert + \vert m+1 \rangle \langle m\vert \right)\right\}$, where the cyclic boundary condition is applied, and the nearest-neighbor coupling, $V$, is negative.
In this ideal J-aggregate, only the optical transition to the lowest exciton state is allowed, and hence the Stokes shift is identical to the reorganization shift.

%absorption and relaxed fluorescence spectra
We calculate the absorption and emission spectra based on the RECKA models.
Essentially, the absorption spectra can be calculated from the Fourier transform of the dipole autocorrelation function~\cite{May2011},
\begin{equation}
I(\omega) \propto Re\int_0^{\infty}dt^{i\omega t} M(t).
\end{equation}
In the Ehrenfest-based method, the dipole autocorrelation function can be obtained from the ensemble average of the time-evolution operator~\cite{Damjanovic2002},
\begin{equation}
\label{eqn:dcor}
M(t) = \left<\sum_{m,n}\mathbf{\mu}_m \mathbf{\mu}_n U_{mn}(t,0)\right>,
\end{equation}
where $U_{mn}$ is the ($m$,$n$)-th element of the time-evolution operator for $H_S+H_{SB}(t)$, and $\mu_m$ denotes a transition dipole moment of $m$-th site. 
Relaxed fluorescence spectra can be obtained from the dipole autocorrelation function averaged with initial conditions in the electronic excited-state~\cite{Lax1952}.
When we propagated wavefunctions and calculated Eq.~\ref{eqn:dcor} for the fluorescence spectra, we sampled $\Delta \epsilon(t=0)$ from a Gaussian distribution of same variance and mean value of $-2\lambda/N$. The lowest exciton state was selected as the initial condition for the exciton propagation.

%Monomer spectra: Mukamel's results, lineshape, SS as a function of kappa

To compare our results with those previously published, we focus on a monomer spectra where analytical solutions are available for the Lorentz-Drude spectral density~\cite{Bosma1990,Mukamel1995}.
For slow relaxation time, the Stokes shift is known to become 2$\lambda$, and the lineshapes become Gaussian. In the opposite limit (i.e., fast relaxation), the absorption and fluorescence spectra become the same without the Stokes shift.
In Fig. 5, we present absorption and fluorescence spectra obtained from the RECKA approach. The lineshape is Gaussian for the fast relaxation time of $\Gamma$ = 10$V$, whereas it becomes Lorentzian for slow relaxation times. To describe the Stokes shifts as a simple function, we introduce the dimensionless parameter $\kappa=\Gamma / \Delta$~\cite{Mukamel1995}. Fig. 6 (a) illustrates the Stokes shifts of the monomer as a function of $\kappa$. For the KA model where the two-time correlation is given by an exponentially decaying function, we have found that the Stokes shifts are exponentially dependent on $\kappa$. We can fit the Stokes shifts to the expression $2\lambda \exp{\left(-\alpha \kappa\right)}$ with $\alpha \cong 0.708$. Obtaining the value of $\alpha$ analytically remains an open problem.
\begin{figure}[!ht]
\begin{center}
 	\includegraphics[width=8cm]{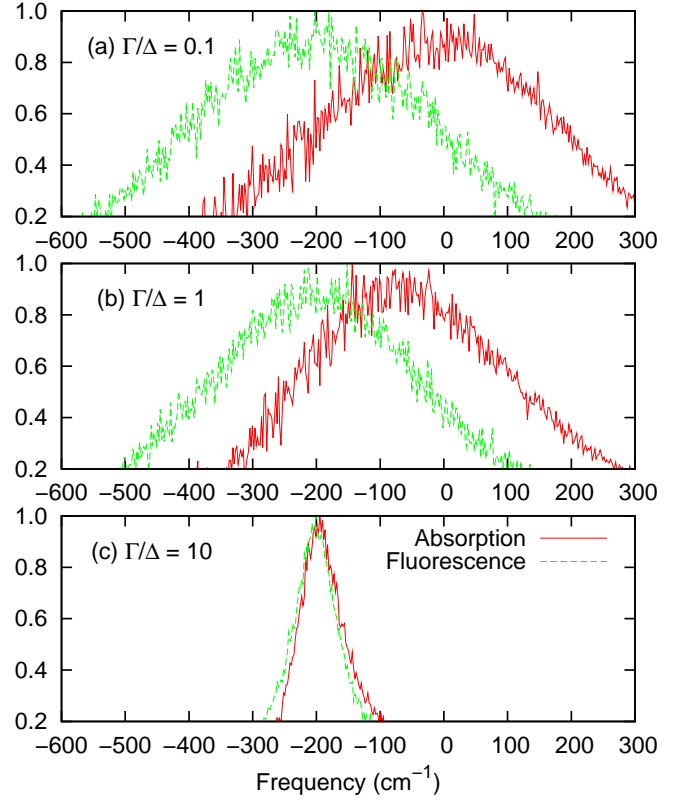}
\end{center}
\caption{Absorption (red-solid) and fluorescence (green-dashed) for the monomer obtained from the RECKA model.}
\end{figure}

\begin{figure}[!ht]
\begin{center}
 	\includegraphics[width=8cm]{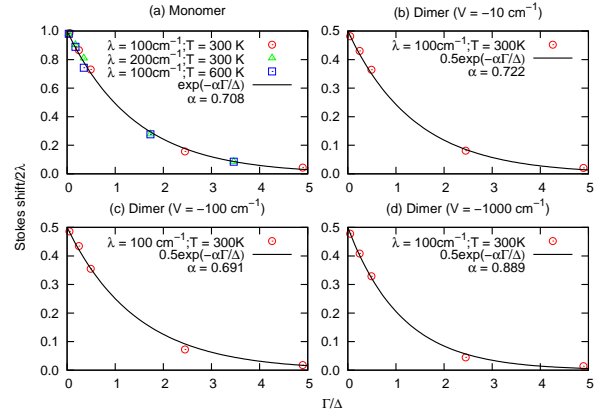}
\end{center}
\label{fig:dsp}
\caption{Stokes shifts as a function of inverse relaxation times ($\Gamma$) for (a) monomer, (b) dimer with the excitonic coupling ($V$) of $-$10 cm$^{-1}$, (c) $-$100 cm$^{-1}$, and $V$=$-$1000 cm$^{-1}$. The Stokes shifts are scaled by twice the reorganization energies ($\lambda$), and the inverse of the relaxation times ($\Gamma$) are scaled by the standard deviations of site energies ($\Delta$).}
\end{figure}

%dimer SS as a function of kappa, different V.
Figs. 6 (b)-(d) illustrates the fits for $\alpha$ for the dimer with $V$=$-$10 , $-$100, and $-$1000 cm$^{-1}$, respectively.
We have obtained the same exponential dependence of $\kappa$, and the Stokes shifts of the dimer become half of those of the monomer. This result suggests that the delocalization of the exciton over the aggregate can lead to smaller reorganization shifts.
Here, $\alpha$ differs slightly for the dimer with different excitonic couplings.

\begin{figure}[!ht]
\begin{center}
 	\includegraphics[width=8cm]{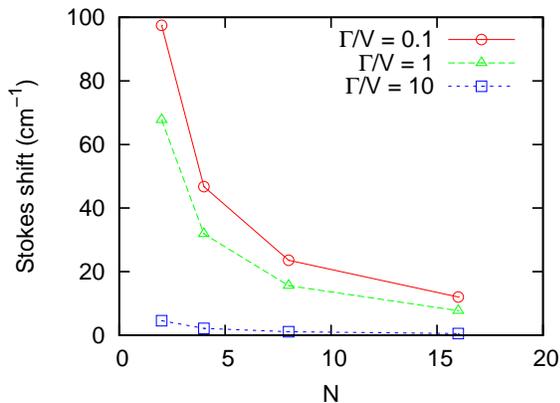}
\end{center}
\caption{Stokes shifts as a function of the number of sites ($N$) obtained with the excitonic coupling of $-$100 cm$^{-1}$ and with different ratios ($\Gamma/V$).}
\end{figure}

Finally, we have obtained the Stokes shifts as a function of the number of sites. As expected from the correction term, $K_m \propto 1/N$; Stokes shifts are inversely proportional to $N$.
From the above results for the Stokes shifts, we have obtained a simple expression for the Stokes shifts for the ideal J-aggregates:
\begin{equation}
\label{eqn:ssf}
\mathrm{Stokes~shift} = \frac{2\lambda}{N}\exp{\left( -\alpha \frac{\Gamma}{\Delta} \right)}.
\end{equation}
Here, $\alpha$ = 0.7 to 0.8.
% Explanation: relationship with experiments 
% Stokes shifts in realistic J-aggregates; reorganization vs internal relxation (between energy level), 3-pulse photon echo (work by T. Goodson)
Stokes shifts attributed to the reorganization vanish when the bath correlation time is fast or when $N$ is large. In macroscopic systems where $N$ is sufficiently large, the reorganization shifts essentially becomes zero. This result strongly suggests that Stokes shifts observed in J-aggregates result from the internal energy relaxation within the exciton manifold. Three-pulse photon echo peak shift experiments~\cite{Donehue2011} have indicated that in cyclic dye aggregates the exciton-phonon coupling decreases as the ring size increases, which is consistent with the present result.
We stress that Eq.~\ref{eqn:ssf} has been obtained by using the ideal J-aggregates, where the exciton can delocalize over the entire aggregate. In a inhomogeneous system exciton delocalization length would be limited by static disorder. In this case, $N$ in Eq.~\ref{eqn:ssf} would be replaced by the inverse participation ratio~\cite{Dahlbom2001,Novoderezhkin2005}.

\section{Conclusion}
In this paper, we have discussed the connection between the atomistic Ehrenfest equation and stochastic bath models.
%By exploiting Mori's projection operator technique, we have shown that the stochastic models such as the HSR can be derived as approximations of the atomistic Ehrenfest equations for exciton-phonon systems.
By exploiting Mori's projection operator technique, we have discussed that stochastic models such as HSR can be derived as approximations of the atomistic Ehrenfest equations for exciton-phonon systems.
Next, we have proposed a reorganization correction scheme to the KA model. The proposed RECKA scheme produces more accurate population dynamics than the KA model, especially in the strong exciton-phonon coupling regime, and provides similar time scales for inter-site energy transfer as compared to the HEOM. The short-time dynamics in the intermediate regime are also more accurate when compared to the KA model. Finally, we have obtained a simple expression for predicting the reorganization shift in ideal J-aggregates. For the stochastic model with exponential correlation functions, the reorganization shifts depend exponentially on the inverse relaxation times. From our findings, the reorganization shifts can be described by three parameters: the monomer reorganization energy, bath relaxation time, and the exciton delocalization length.  This simple relationship allows the physical origin of the Stokes shift to be understood and will be useful to separate the reorganization shift from internal energy relaxation.

\section*{Acknowledgements}
The authors would like to thank Dr. Semion K. Saikin and Nicolas Sawaya for fruitful discussion. T.F. thanks Dr. Christoph Kreisbeck for providing the HEOM results. T. F., J. H., and A. A.-G. acknowledge support from the Center for Excitonics, an Energy Frontier Research Center funded by the US Department of Energy, Office of Science and Office of Basic Energy Sciences under award DE-SC0001088.
A. A.-G. also acknowledge Defense Threat Reduction Agency grant HDTRA1-10-1-0046 and Defense Advanced Research Projects Agency grant N66001-10-1-4063, as well as the Corning foundation.

\end{document}